\documentclass[cameraready]{Interspeech}

\title{MMGenre: Benchmarking Singing Voice Synthesis \\ across Multiple Musical Genres}

\author[
affiliation={1},
orcid=0009-0005-6970-7439
]{Wenhao}{Feng}

\author[
affiliation={1},
orcid=0009-0002-4538-7440
]{Yuxun}{Tang}

\author[
affiliation={2},
orcid=0000-0002-9050-8304
]{Jiatong}{Shi}

\author[
affiliation={1},
orcid=0000-0001-6486-6020,
correspondingauthor
]{Qin}{Jin}

\address{
$^1$ AIM3 Lab, Renmin University of China, China \\
$^2$ Carnegie Mellon University, United States
}

\email{
wenhaofeng@ruc.edu.cn,
tangyuxun@ruc.edu.cn,
jiatongs@cs.cmu.edu,
qjin@ruc.edu.cn
}


\usepackage{comment}

\usepackage{float}
\usepackage{tabularx}
\usepackage{comment} 
\usepackage{subcaption}
\usepackage{cite}

\newcommand{\benchmarkname}{MMGenre}

\begin{document}

\maketitle





\begin{abstract}
Singing voice synthesis (SVS) has progressed rapidly, yet its ability to generalize across diverse musical genres remains underexplored. Existing benchmarks are heavily biased toward pop music, limiting systematic analysis of genre-dependent behavior.
We introduce MMGenre, a benchmark for multi-genre SVS diagnosis, supported by an automatic pipeline for constructing genre-aligned music scores. MMGenre spans 10 major genres and 26 subgenres, enabling comprehensive analysis of genre-aware synthesis.
Extensive evaluation of representative SVS models reveals limited genre discrimination: synthesized vocals across genres exhibit highly similar acoustic characteristics and weak separability. While zero-shot genre adaptation yields only marginal improvements, lightweight genre-specific continued training leads to substantial gains.
MMGenre provides a standardized framework for multi-genre SVS evaluation and exposes critical challenges in achieving genre-aware singing voice synthesis.


\end{abstract}

\section{Introduction}
\label{sec:intro}


{Singing voice synthesis (SVS)} aims to generate expressive singing voices directly from symbolic music scores and has achieved substantial progress with recent neural and diffusion-based models~\cite{shi22muskits, shi2021sequence, zhang2022visinger, liu2022diffsinger}. 
Prior research has largely focused on improving naturalness~\cite{lu2020xiaoicesing, zhang2022visinger2, wu2024toksing}, expressiveness~\cite{dai2024expressivesinger}, and controllability~\cite{zhang2024stylesinger, guo2025techsinger} of synthesized singing voices.

However, singing performance is inherently shaped by musical genre, a high-level semantic attribute that is immediately recognizable to human listeners. Genres such as pop, rock, jazz, and classical music exhibit systematic differences in vocal timbre, articulation, phrasing, rhythmic emphasis, and expressive conventions~\cite{sundberg1990science}. 
In real-world music production and consumption, genres form stable perceptual categories that shape listener evaluation and preference~\cite{rentfrow2003re, kowald2020utilizing}. Despite its perceptual salience and practical importance, genre has rarely been treated as a first-class evaluation dimension in SVS research. A fundamental yet underexplored question remains: \textit{how well do current SVS models generalize across musical genres?}

This gap is largely due to data and evaluation limitations. Public SVS datasets are overwhelmingly dominated by pop music (e.g., M4Singer~\cite{zhang2022m4singer}, Opencpop~\cite{wang2022opencpop}, ACE-Opencpop~\cite{shi2024aceopencpop}), offering limited coverage of other genres and making large-scale systematic genre-level analysis infeasible. 
Existing style-aware SVS studies primarily target lower-level attributes, such as emotion, tempo, pitch range, or specific singing techniques~\cite{zhang2024stylesinger, zhang2024tcsinger, guo2025techsinger, zhang2025tcsinger2}, rather than modeling genre as a holistic musical condition. Consequently, it remains unclear whether current SVS systems truly capture genre-specific characteristics or simply reproduce surface-level acoustic patterns learned from biased data.

To overcome these data limitations, we turn to recent advances in text-to-music (T2M) generation~\cite{yuan2025yue, ning2025diffrhythm, chen2025diffrhythm+, liu2025songgen, karystinaios2025weavemuse}. Modern T2M systems enable explicit genre specification through natural language prompts. Compared to real-world recordings, T2M-generated music is easier to scale, poses fewer copyright constraints, and can be systematically diversified across genres, making it a practical and scalable resource for constructing genre-balanced evaluation data.

Leveraging these advances, we propose an automatic pipeline that uses T2M models to generate genre-aligned music scores, enabling scalable construction of multi-genre data. Based on this pipeline, we introduce \textbf{MMGenre}, a benchmark for multi-genre SVS evaluation spanning 10 major genres and 26 fine-grained subgenres. MMGenre significantly expands genre coverage beyond existing SVS benchmarks and supports systematic analysis of genre-dependent synthesis behavior. 

Using MMGenre, we evaluate representative SVS models and obtain several key findings. First, synthesized vocals across genres remain acoustically highly similar, even when conditioned on genre-specific scores. Second, inference-time zero-shot strategies (e.g., style transfer or technique conditioning) yield only marginal improvements in genre alignment. In contrast, even limited genre-specific fine-tuning consistently produces substantial gains. These results suggest that genre awareness in SVS is not an emergent, easily transferable capability, but remains strongly tied to training data distribution.

Overall, MMGenre establishes the first systematic framework for evaluating genre generalization in SVS and exposes fundamental challenges in genre-aware modeling. Our contributions are threefold: 
(1) We propose an automatic and scalable pipeline for generating genre-aligned music scores, providing an extensible framework for constructing controlled multi-genre SVS evaluation data.~(2) We introduce {\benchmarkname}, the first benchmark for systematic evaluation of SVS across diverse musical genres.~(3) We provide a comprehensive empirical study of training- and inference-time strategies for improving genre alignment, revealing the data dependence and limitations of current SVS models.
We hope MMGenre will facilitate multi-genre evaluation and
encourage genre-aware SVS research. Benchmark resources,
including dataset access, source code, and audio demonstrations,
are available through the project page~\footnote{\url{https://fengjin1117.github.io/mmgenre-web/}}.




\section{Multi-Genre Benchmark}
\label{sec:benchmark}
\subsection{Benchmark Overview}
\begin{figure}[!t]
    \centering
    \includegraphics[
      width=0.85\linewidth,
      trim=45 45 45 45,
      clip
    ]{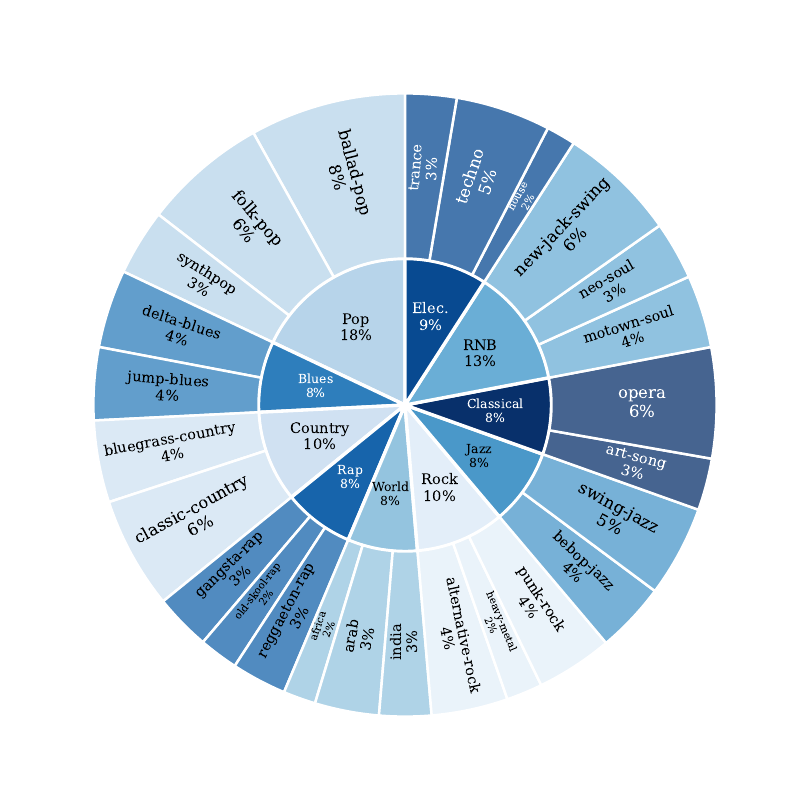}
    \vspace{-6pt}
    \caption{Hierarchical distribution of genres (inner ring) and subgenres (outer ring) in the MMGenre benchmark, with proportions computed based on segment counts.}
    \label{fig:genre_subgenre_sunburst}
    \vspace{-6pt}
\end{figure}

MMGenre is designed to provide a unified benchmark for evaluating SVS under diverse musical genre settings. 
The benchmark covers 10 major musical genres and 26 representative subgenres organized in a hierarchical taxonomy. 
As illustrated in Fig.~\ref{fig:genre_subgenre_sunburst}, each major genre is associated with 2–3 representative subgenres to ensure stylistic diversity and reduce pop-dominant bias.

MMGenre consists of 3,152 aligned Chinese score–audio segment pairs, with a total duration of approximately 4.36 hours. 
Each segment is constrained to 2–8 seconds, with an average duration of approximately 5 seconds, ensuring compatibility with most SVS inference settings while preserving expressive singing characteristics.

\subsection{Data Construction Pipeline}

\begin{figure}[t]
    \centering
    \includegraphics[width=0.85\linewidth]{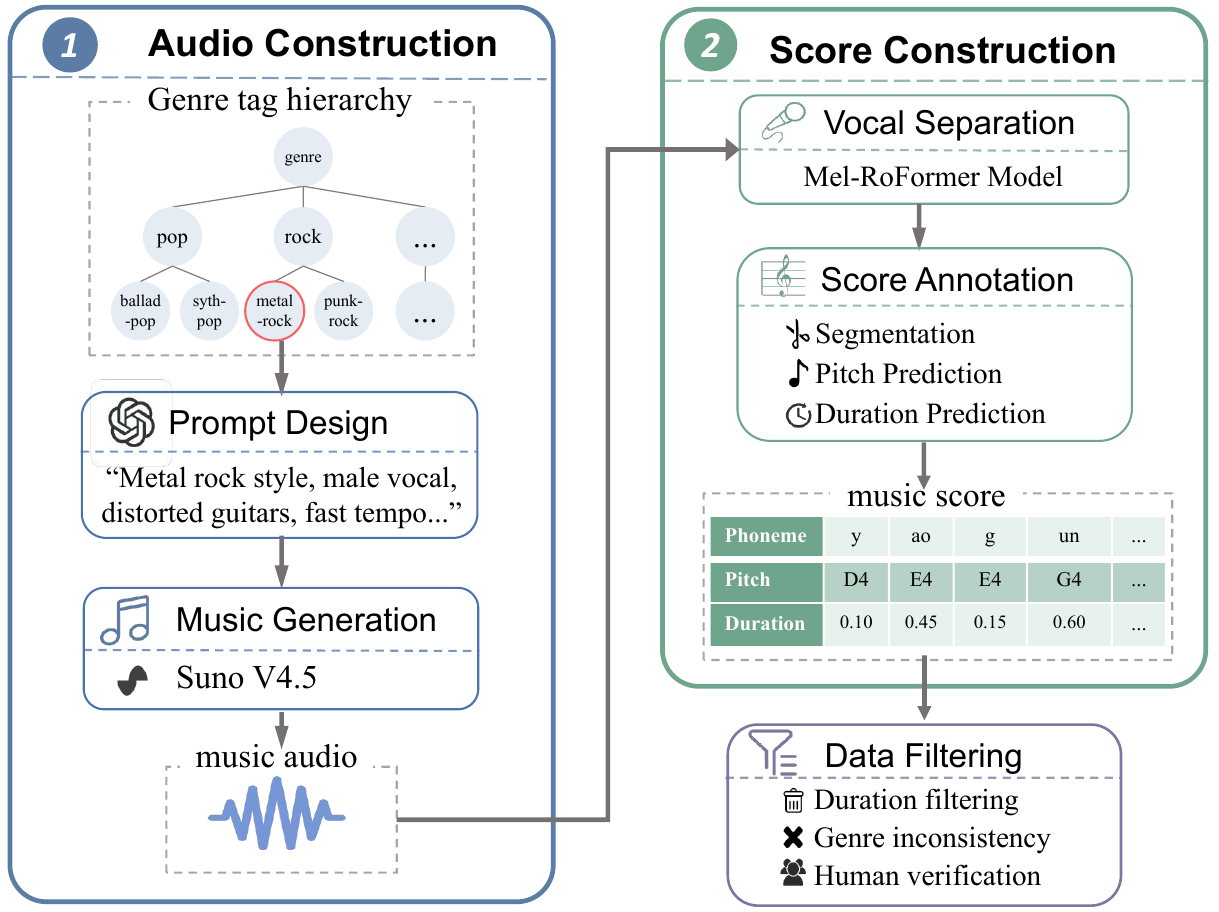}
    \vspace{-6pt}
    \caption{Benchmark data construction pipeline.
    The final benchmark data consist of genre-conditioned singing voice and aligned symbolic music score pairs.}
    \label{fig:data_construction}
    \vspace{-6pt}
\end{figure}

As shown in Fig.~\ref{fig:data_construction}, we construct MMGenre through a unified and scalable pipeline that converts genre-conditioned music into aligned singing voice--score pairs. 
Starting from a hierarchical genre taxonomy, we first employ a text-to-music system to synthesize raw genre-conditioned music. 
Next, the generated audio is processed through a score construction stage, where phoneme-aligned symbolic scores are derived. 
The original full-mix music is retained as auxiliary ground truth for genre style verification, while the final benchmark data consist of disentangled singing voice and corresponding score pairs.



\noindent\textbf{Audio construction.}
We employ Suno V4.5~\cite{Suno} as the text-to-music generator within our pipeline to produce genre-conditioned music samples via textual prompts.
Prompt design follows a hierarchical genre taxonomy and is initially assisted by ChatGPT~\cite{achiam2023gpt}, followed by manual refinement to ensure clarity and stylistic relevance. 

\noindent\textbf{Score construction.}
Generated music samples are separated into vocal tracks using Mel-RoFormer~\cite{Wang2023MelBandRF}. We then apply STARS~\cite{guo2025stars} for phoneme-level pitch and duration annotation, producing standardized symbolic music scores aligned with singing audio.

\noindent\textbf{Data filtering.}
The generated audio–score pairs are further processed through a data filtering stage to ensure reliability. Segments with abnormal durations are removed, and genre consistency is automatically verified using MuQ-MuLan~\cite{zhu2025muq}. A small portion of samples is then checked by human listeners to eliminate occasional artifacts introduced by automatic processing. The resulting singing voice–score pairs are used as standardized inputs for SVS inference in all benchmark experiments.

Beyond supporting this benchmark, the pipeline itself provides a reusable framework for constructing genre-balanced SVS data under controlled and reproducible conditions. The modular design of the pipeline allows straightforward extension to new genres, languages, or alternative text-to-music generators without modifying the downstream evaluation protocol.

%


\subsection{Diversity Analysis}



\begin{figure}[!t]
    \centering
    \begin{subfigure}[t]{0.43\linewidth}
        \centering
        \includegraphics[width=\linewidth]{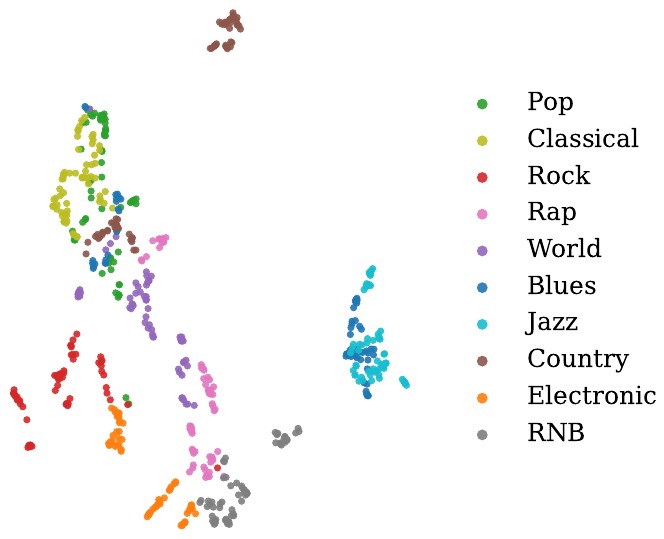}
        \caption{Audio embedding space}
    \end{subfigure}
    \hfill
    \begin{subfigure}[t]{0.55\linewidth}
        \centering
        \includegraphics[width=\linewidth]{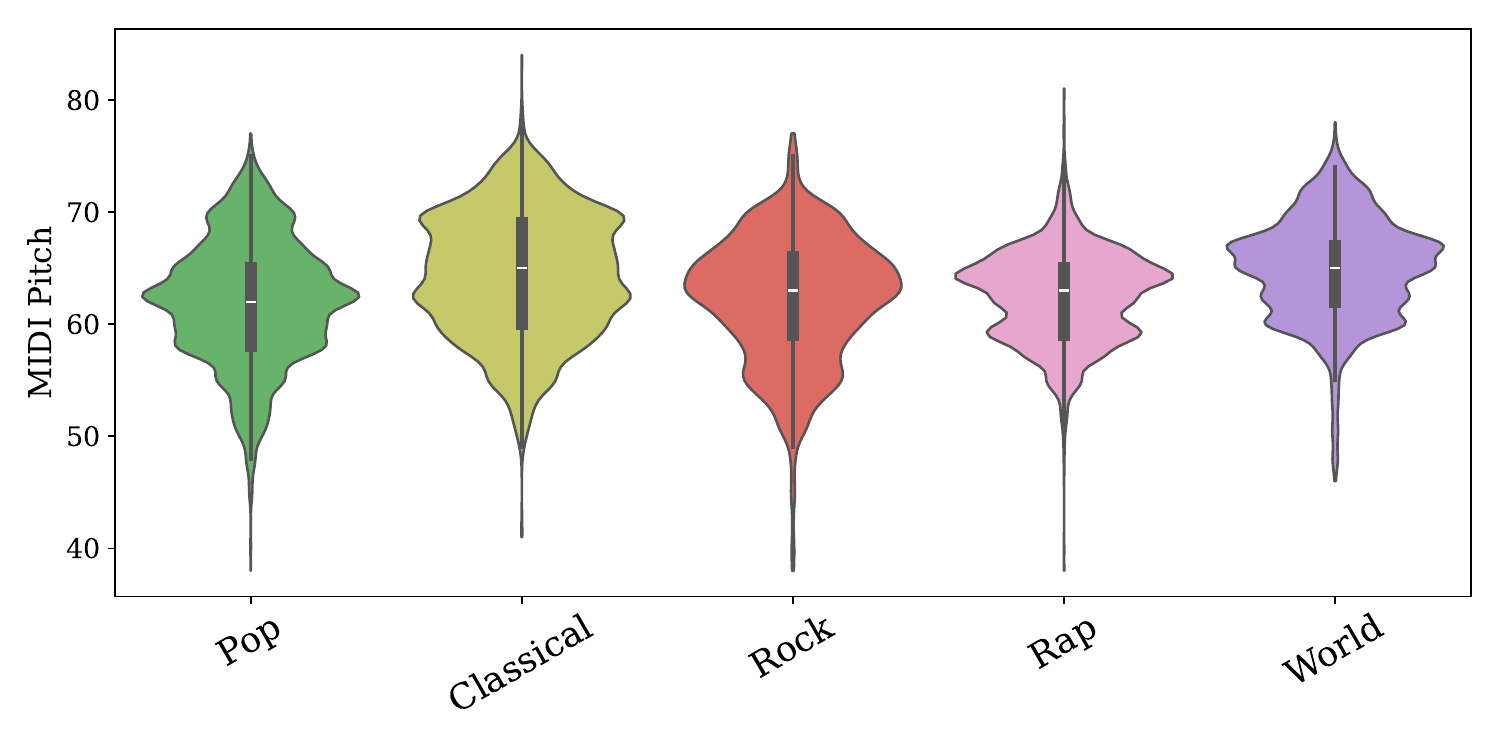}
        \caption{Score-level pitch distribution}
    \end{subfigure}
    
    \caption{
    \textbf{Diversity analysis from audio and symbolic perspectives.}
    (a) UMAP visualization of MuQ audio embeddings for genre-conditioned Suno-generated samples.
    (b) Distribution of MIDI note pitches across selected genres.
    }
    \label{fig:gt_diversity_joint}
\end{figure}

\noindent\textbf{Audio-level diversity.}
We visualize MuQ audio embeddings of genre-conditioned Suno samples using UMAP, as shown in Fig.~\ref{fig:gt_diversity_joint}(a). Distinct clusters emerge across genres, and MuQ-MuLan achieves strong recognition accuracy (Top-1: 76\%, Top-3: 92\%), confirming reliable cross-genre separability.


\noindent\textbf{Score-level diversity.}
Beyond acoustic embeddings, we further analyze symbolic score statistics by examining the distribution of MIDI note pitches across genres, as shown in Fig.~\ref{fig:gt_diversity_joint} (b). 
These systematic differences in pitch range and dispersion demonstrate structural diversity across genres at the score level.

\subsection{Benchmark Validity Analysis}



To assess the impact of singing data distribution on benchmark reliability, we investigate whether Suno-synthesized singing affects the relative evaluation of SVS systems. To this end, we conduct a controlled human MOS study comparing model performance on real and synthesized singing in the Pop genre. To minimize confounding factors, validation is restricted to Pop, aligning with existing SVS datasets. Both conditions share identical musical scores constructed through the same pipeline, differing only in singing source. Eight SVS models are evaluated, with 35 samples per condition (real and synthesized) rated by eight annotators (4,480 ratings in total).

The relative ranking of SVS models remains highly consistent across the two conditions. The Spearman rank correlation between MOS rankings on real and synthesized data reaches $\rho = 0.90$ ($p < 0.01$), indicating strong agreement in model ordering. These findings demonstrate that although the singing source differs, synthesized singing preserves the relative comparative relationships among SVS systems.

\section{Experiments}
\label{sec:experiements}

The central question of this study is \textit{whether current SVS models can generate genre-consistent singing voices from genre-specific musical scores?}
To systematically investigate this issue, we conduct a structured empirical analysis that examines genre behavior from perceptual, representational, and training-dynamics perspectives.



\subsection{Experimental Setup}
\subsubsection{Models}

We evaluate a diverse set of representative singing voice synthesis (SVS) models covering both autoregressive and non-autoregressive paradigms. Specifically, RNN~\cite{shi2021sequence}, \mbox{XiaoiceSing}~\cite{lu2020xiaoicesing}, VISinger~\cite{zhang2022visinger}, and VISinger2~\cite{zhang2022visinger2} are trained using the official implementations provided by the ESPnet~\cite{watanabe2018espnet} toolkit, with Opencpop~\cite{wang2022opencpop} as the training dataset. In addition, DiffSinger~\cite{liu2022diffsinger}, StyleSinger~\cite{zhang2024stylesinger}, TCSinger~\cite{zhang2024tcsinger}, and TechSinger~\cite{guo2025techsinger} are evaluated using their publicly released pretrained models from GitHub. 

\subsubsection{Evaluation Metrics}

\noindent\textbf{Genre Alignment Metrics.}
Genre Consistency Score (GCS-5) is the core evaluation metric in this work, defined as a 5-point perceptual rating that measures how well synthesized singing matches the target genre.
Following recent studies that employ large multimodal models for audio-related evaluation~\cite{lin2025voice, lee2025performance, manku2026emergenttts}, we adopt Gemini~2.5~Pro~\cite{comanici2025gemini} as an automatic rater.
Given a standardized prompt, Gemini~2.5~Pro assigns a genre consistency score on a 5-point Likert scale, where higher scores indicate better alignment.
Unless otherwise specified, all genre-related analyses in this paper are based on GCS-5.
A human agreement check on 100 benchmark samples
spanning five representative genres shows strong correlation
between Gemini and human ratings (Spearman $\rho = 0.85$,
$p < 0.01$), supporting the use of GCS-5 as a scalable proxy
for analyzing relative genre-alignment trends.



\noindent\textbf{Singing Quality Metrics.}
We report pseudo-MOS predictors (SingMOS~\cite{tang2024singmos}, SingMOS-Pro~\cite{tang2025singmospro}, and SSQA~\cite{huang2024ssqa}) together with a character error rate (CER) metric to measure 
lyric intelligibility. 
CER is computed using the Whisper-WER implementation in VERSA~\cite{shi2025versa}. 
Those metrics are used to ensure that observed genre-related effects are not caused by degraded synthesis quality. 

\subsection{Benchmark Results}

\subsubsection{Genre-wise Alignment Results}
\label{subsec: genre_eval}

\begin{figure}[!t]
    \centering
    \includegraphics[width=0.88\columnwidth]{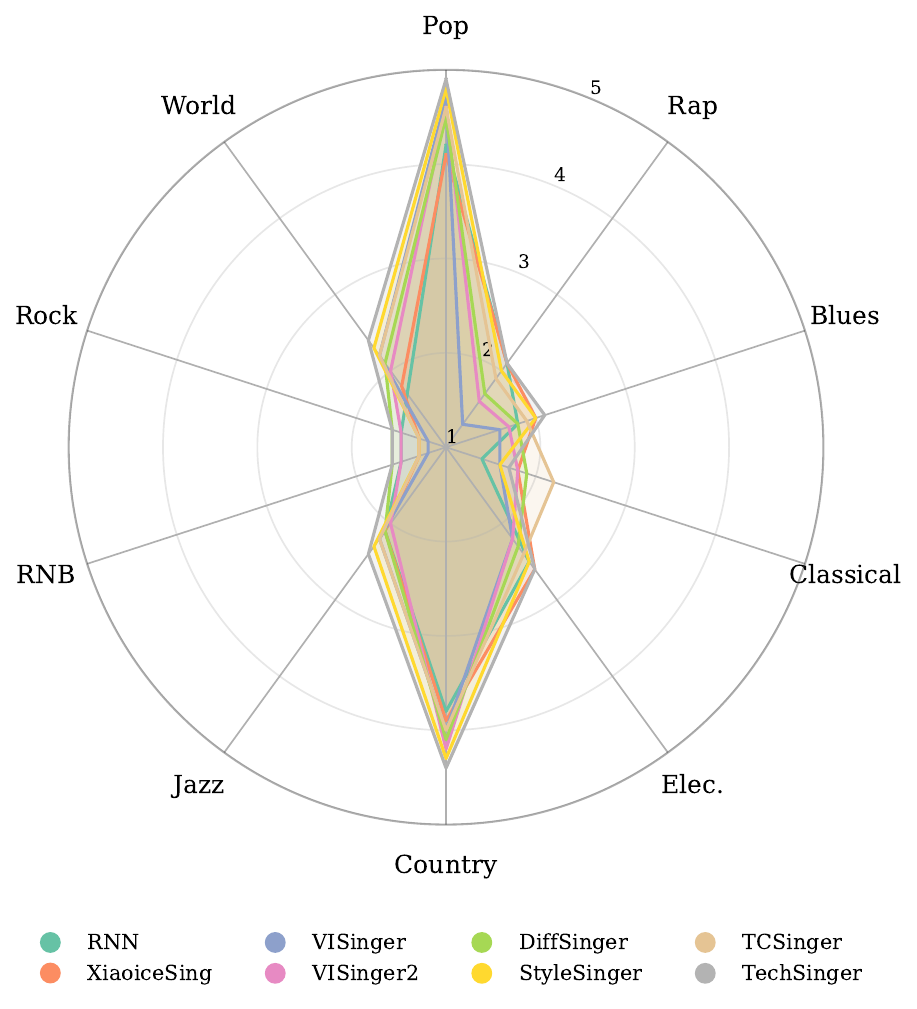}
    \caption{Genre-wise GCS-5 radar plot across SVS models. 
    All models exhibit highly similar genre profiles, with strong alignment concentrated in Pop-related genres, revealing a systematic genre collapse phenomenon.}
    \label{fig:genre_align_radar}
\end{figure}

\begin{figure}[t]
    \centering
    \begin{minipage}{0.48\columnwidth}
        \centering
        \includegraphics[width=\linewidth]{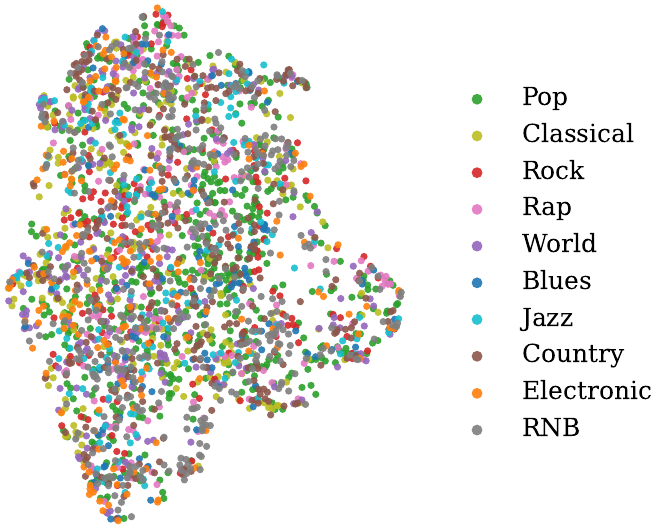}
        \vspace{2pt}
        {\small (a)}
    \end{minipage}
    \hfill
    \begin{minipage}{0.48\columnwidth}
        \centering
        \includegraphics[width=\linewidth]{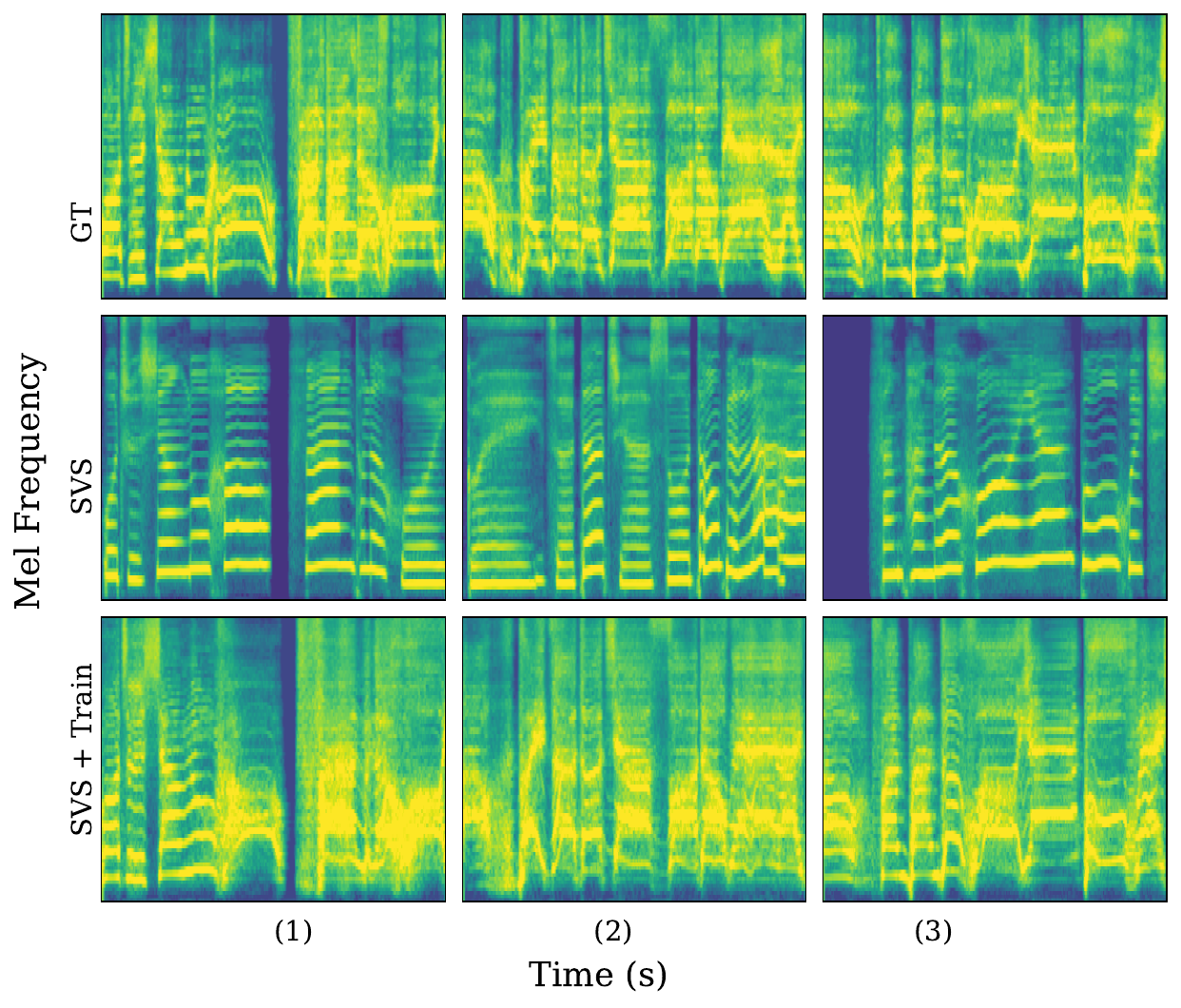}
        \vspace{2pt}
        {\small (b)}
    \end{minipage}
    \caption{
    Visualization of the genre collapse phenomenon.
    (a) UMAP projection of MuQ embeddings for SVS inference audio, showing substantial cross-genre overlap.
    (b) Log-mel spectrogram comparison on Rock  across three representative samples (columns 1–3).
    }
    \label{fig:genre_collapse_visual}
\end{figure}

We report genre alignment metrics of different SVS systems across all genre categories. 
As shown in Fig.~\ref{fig:genre_align_radar}, high genre alignment is consistently observed only for Pop and a limited subset of acoustically related genres, while most non-Pop genres receive uniformly low scores across models. 
This trend is consistently reflected in the embedding space. The UMAP visualization in 
Fig.~\ref{fig:genre_collapse_visual} (a) 
shows that synthesized samples from different genres form highly overlapping distributions in the embedding space, in contrast to the clearly separable structure observed for ground-truth data. This indicates a relative collapse of genre-discriminative acoustic features after synthesis.

Most out-of-distribution genres with vocal styles that differ substantially from Pop, such as Rock, Rap, and Classical, exhibit consistently low alignment scores across architectures. 
In contrast, certain unseen genres with vocal characteristics closer to Pop, such as Country, achieve comparatively higher alignment. 
This pattern suggests that, under the controlled benchmark setting, current SVS systems tend to reproduce dominant vocal style priors learned from training data rather than consistently encoding genre-specific attributes as an independent controllable factor. 


To further examine the genre collapse phenomenon, 
Fig.~\ref{fig:genre_collapse_visual} (b)
presents log-mel spectrogram comparisons on the Rock genre, using punk rock as a representative subgenre. 
The ground-truth recordings exhibit broadband high-frequency energy, sharp transient structures, and irregular harmonic spreads characteristic of punk rock vocals. In contrast, the SVS outputs display smoother harmonic contours and more regular time–frequency patterns, with attenuated high-frequency components. Such differences suggest that genre-specific acoustic cues are not faithfully preserved in the synthesized results.



\subsubsection{Overall Synthesis Quality}
\label{subsec: overall_results}



Table~\ref{tab:overall_evaluation} reports overall subjective 
quality prediction and intelligibility results across all compared 
SVS models, computed as macro-averages over genres to ensure 
equal contribution. As expected, earlier autoregressive baselines 
obtain lower scores, while more recent SVS systems achieve higher 
performance across both pseudo-MOS predictors and CER, despite the speech-centric nature of ASR systems, reflecting general progress in synthesis quality and lyric clarity. 

Notably, this performance trend remains consistent even under the 
newly constructed multi-genre distribution, suggesting that overall 
synthesis quality improvements generalize across diverse genre conditions.



\begin{table}[!t]
\centering
\footnotesize
\caption{
Overall SVS evaluation results on MMGenre, aggregated over all genres.
SingPro denotes SingMOS-Pro.
}
\begin{tabular}{l@{\hspace{4pt}}cccc}
\toprule
\textbf{Model} 
& \textbf{SingMOS↑} 
& \textbf{SingPro↑} 
& \textbf{SSQA↑} 
& \textbf{CER↓} \\
\midrule
RNN~\cite{shi2021sequence}         
& 3.25 & 2.94 & 2.87 & 0.47 \\
XiaoiceSing~\cite{lu2020xiaoicesing}  
& 3.43 & 3.08 & 2.97 & 0.69 \\
TechSinger~\cite{guo2025techsinger}  
& 3.84 & 3.76 & 3.78 & 0.59 \\
StyleSinger~\cite{zhang2024stylesinger} 
& 3.98 & 3.74 & 3.72 & 0.54 \\
TCSinger~\cite{zhang2024tcsinger}   
& 3.98 & 3.84 & 3.85 & 0.65 \\
VISinger~\cite{zhang2022visinger}    
& 3.94 & 3.92 & 4.00 & \textbf{0.25}\\
VISinger2~\cite{zhang2022visinger2}  
& 3.97 & 3.98 & \textbf{4.07} & \textbf{0.25} \\
DiffSinger~\cite{liu2022diffsinger}  
& \textbf{4.08} & \textbf{4.04} & 4.06 & 0.31 \\
\bottomrule
\end{tabular}
\label{tab:overall_evaluation}
\end{table}

\subsection{Mechanistic Analysis of Genre Collapse}
\label{subsec: inference_and_training}

The observed genre collapse raises a key question: 
does the failure originate from insufficient genre-discriminative information in the symbolic score, or from dominance of the training data distribution?
To disentangle them, we analyze the problem from two perspectives: 
(1) \textit{symbolic controllability at inference time}, 
and~(2) \textit{distributional bias during training}. 
This analysis reveals whether genre behavior can be activated through score-level conditioning alone, or whether it is primarily determined by learned data priors.

\begin{table}[!t]
\centering
\small
\caption{GCS-5 under zero-shot genre control.
Improvements from score-level or inference-time conditioning remain limited.}
\begin{tabular}{lccc}
\toprule
Method & Classical & Rock & Rap \\
\midrule
StyleSinger                     & 1.6 & 1.3 & 1.4 \\
StyleSinger + Style Transfer    & 2.6 & 1.7 & 1.6 \\
TechSinger                      & 1.7 & 1.6 & 1.5 \\
TechSinger + Technique Control  & 2.1 & 1.7 & 1.5 \\
\midrule
Ground Truth                       & 4.2 & 4.8 & 4.4 \\
\bottomrule
\end{tabular}
\label{tab:zeroshot_genre_control}
\vspace{-6pt}
\end{table}

\noindent\textbf{Inference-Time Symbolic Controllability.}
If conventional phoneme–pitch–duration score representations encode sufficient genre-discriminative cues, 
then inference-time control mechanisms should be able to activate distinct vocal realizations without modifying model parameters. 
To test this hypothesis, we evaluate two zero-shot control strategies available among the evaluated SVS systems, namely reference-based style transfer (StyleSinger) and phoneme-level technique annotation (TechSinger), as shown in Table~\ref{tab:zeroshot_genre_control}.

Although minor improvements are observed in certain cases, 
alignment scores remain substantially below those of Suno GT. 
More importantly, the relative gaps across genres persist, 
and non-Pop categories continue to exhibit weak differentiation. 
These results suggest that the standard symbolic score representation 
provides limited genre-specific information, 
restricting controllability at inference time. 
In other words, genre awareness does not emerge 
simply by augmenting inference-time conditioning 
when the underlying representation remains unchanged.

\noindent\textbf{Training-Time Distributional Bias.}
We next examine whether genre collapse is primarily governed 
by training data statistics. 
To this end, we conduct genre-specific continued training on Rock 
using two hours of independently constructed AI-generated data, disjoint from the benchmark set.
Under this setting, the GCS-5 score improves dramatically 
from 1.5 to 4.9. 
The slightly higher score than GT may reflect subgenre-specific exaggeration of salient Rock cues (e.g., metal or punk characteristics), which can increase perceived genre strength under GCS-5.
As shown in Fig.~\ref{fig:genre_collapse_visual}(b), 
the fine-tuned model exhibits stronger high-frequency energy 
and sharper transient structures, 
closer to ground-truth Rock vocals, 
while the baseline output retains smoother, Pop-like characteristics.

The substantial improvement indicates that genre-consistent vocals can emerge with sufficient genre-specific training data. This contrast with the limited zero-shot gains demonstrates that genre behavior in current SVS systems is primarily governed by learned distributional priors rather than symbolic controllability.

\noindent\textbf{Implications.}
Taken together, these findings suggest that 
genre awareness in existing SVS models 
is not an emergent property of score-level conditioning, 
but a distribution-dependent behavior shaped by training data composition. 
Overall quality metrics may therefore obscure systematic stylistic collapse, 
highlighting the necessity of genre-aware benchmarks 
for diagnosing data-driven biases in singing voice synthesis.

\section{Conclusion}
\label{sec:conclusion}
We present \benchmarkname, a unified benchmark for evaluating genre-aware singing voice synthesis across multiple musical genres under standardized and controlled conditions. 
Through systematic evaluation, we reveal a consistent genre collapse phenomenon: current SVS systems exhibit limited genre differentiation and tend to converge toward a dominant Pop-like style. 
We further show that inference-time control yields only marginal gains, while even limited genre-specific continued training substantially improves alignment, highlighting the dominant role of training data distribution. 
We hope \benchmarkname\ will facilitate standardized evaluation of genre-aware SVS.

\section{Acknowledgments}
This work was partially supported by the National Natural Science Foundation of China (No. 62576347). We thank Renke Huang for insightful discussions and suggestions on music-related aspects of this work, particularly on the design of the genre hierarchy.

\section{Generative AI Use Disclosure}
Generative AI tools were used solely for language polishing and improving the clarity of expression. All technical content, experimental design, results, and conclusions were produced and verified by the authors.

\bibliographystyle{IEEEtran}
\bibliography{mybib}

\end{document}